\documentclass[sigconf,authorversion]{acmart2}

\usepackage{xcolor}

\usepackage{dirtytalk}
\usepackage{multirow}
\usepackage{float}

\usepackage{macros}
\usepackage{xcolor}
\colorlet{BLACK}{black}
\newcommand{\change}[1]{{\leavevmode\color{black}{#1}}}

\AtBeginDocument{%
  \providecommand\BibTeX{{%
    \normalfont B\kern-.5em{\scshape i\kern-.25em b}\kern-.8em\TeX}}}

\setcopyright{rightsretained}
\acmConference[CHI '23 Workshop]{2023 CHI Conference on Human Factors in Computing Systems: Integrating Individual and Social Contexts into Self-Reflection Technologies Workshop}{April 23, 2023}{Hamburg, Germany}
\acmBooktitle{2023 CHI Conference on Human Factors in Computing Systems: Integrating Individual and Social Contexts into Self-Reflection Technologies Workshop, April 28, 2023, Hamburg, Germany}

\usepackage{array}
\begin{document}

\newcommand{\PreserveBackslash}[1]{\let\temp=\\#1\let\\=\temp}
\newcolumntype{C}[1]{>{\PreserveBackslash\centering}p{#1}}
\newcolumntype{R}[1]{>{\PreserveBackslash\raggedleft}p{#1}}
\newcolumntype{L}[1]{>{\PreserveBackslash\raggedright}p{#1}}

\title[]{Value Reflection: Motivation, Examples and The Role of Technology}


\author{Ken Jen Lee and Edith Law}
\email{{kenjen.lee, edith.law}@uwaterloo.ca}
\affiliation{%
  \institution{University of Waterloo}
  \streetaddress{200 University Avenue West}
  \city{Waterloo}
  \state{Ontario}
  \country{Canada}
  \postcode{N2L 3G1}
}

\renewcommand{\shortauthors}{Lee and Law}

\begin{abstract}
Findings from existing research provide possible evidence that values are an important construct that should be part of the self-reflection process.
However, many questions about value reflection remain unexplored.
As such, this position paper aims to provide an overview of relevant research and frameworks,
example studies on value reflection pursued by the authors, and design aspects that might both improve value reflection processes and provide a brief mapping of where and how technology could be of help.

\end{abstract}


\maketitle

\section{Why Reflect on Values?}
Value reflection refers to the process of reflecting on an individual's own values.
While it has been explored in existing research to some degree (e.g., \cite{huldtgren2014designing, pommeranz2012elicitation}), many open questions about its implementation, effects, and potential avenues for technological support are still unanswered. 
However, existing research in the social sciences provide strong reasons for HCI researchers to pursue further investigations on value reflection technologies. 
One such reason is values' interesting connection to other constructs that could predict people's behaviors. 
Compared to interests and attitudes, which are judgement based on liking, value judgements are based on importance \change{\cite{dawis1991vocational,brown2004career,DAWIS200116299}}. Moreover, both interests and attitudes could serve as expressions of values, i.e., reflecting on values is a way of reflecting on reasons behind an individual's interests and attitudes. 
Similarly, values act as a mechanism that activates context-specific goals that align with an individual's values, and as a guidance for an individual's actions towards those goals \change{\cite{maio2016psychology,lewin1951field}}. In other words, reflecting on values can improve people's understanding on how they activate goals and guide behaviors.
Compared to personality traits, researchers found values to have ``more conceptual potential for within-person change'' \cite{maio2016psychology}. 
Moreover, social sciences research also found that an individual's values could be influenced by various personal and social factors \cite{maio2016psychology}. Example of important personal factors include the values of their family and/or caregivers \change{\cite{roest2009value,garnier1998values,hoge1982transmission}}, age \change{\cite{glenn1980values,nunn1978tolerance,sheldon2001getting}} and personal biology \change{\cite{brosch2012importance,brosch2011generating}}. Example of important social factors include education \change{\cite{tal2002teachers,assor1997value}}, culture \change{\cite{markus1991culture,bochner1994cross,triandis2018individualism}}, mass media consumption \change{\cite{dittmar2007consumer,twenge2010birth}}, and socioeconomic status \change{\cite{van2012current}}. 
As such, reflecting on how an individual's values are formed, and values' effects, can naturally open the door for deeper reflections on an individual's personal and social factors, and how their values inform their behaviors and ways of thinking.

As Le Dantec et al. put it, there is an ``inherent difficulty in talking about values'' \cite{le2009values}. To keep discussions at a high-level, this position paper will not adhere to a single definition of what values are, neither will it adhere to a certain value framework (works like \cite{maio2016psychology} discuss these in detail). Determining the definition of values and which framework to use is very likely not universal and context dependent. However, the definition of human values in the value-sensitive design paradigm as ``what is important to people in their lives, with a focus on ethics and morality'' \cite{friedman2019value} provides a good overview of the kind of values we mean to discuss, i.e., not ``how much something is worth in money'' \cite{valueoxford}. \change{Specifically, in Section \ref{sec:relwork}, we discuss exisiting works on value reflection and relevant research on Value-Sensitive Design and reflection frameworks. Then, we discuss elements that could be crucial to the value reflection process in Section \ref{sec:designaspects}. Lastly, we provide two examples of how value reflection could be designed in different contexts (Section \ref{sec:examples}). }

\section{Relevant Works}\label{sec:relwork}
\subsection{Value Reflection}
In \cite{huldtgren2014designing}, Huldtgren et al. explored technologically supported value reflection towards supporting users in making better life decisions by better considering their values. They used a website-based technology probe where participants could reflect on various content types (e.g., stories, photos) by describing the content and selecting a related value from a predefined list of values. The reflection flow is designed to guide participants from concrete (content description) to abstract (value reflection) tasks. The website also provides participants with a tag cloud showing the frequency of selected values. Through this study, many important aspects of value reflection, which we discuss below, were brought up as well.
Pommeranz et al. \cite{pommeranz2012elicitation}, on the other hand, explored how technology could support the process of reflecting on context-specific values (which they refer to as situational values) in-situ using the photo elicitation method through a mobile application.
Specifically, participants were asked to take 10 pictures and tag them with values using the mobile application. Then, they were interviewed on why the photos were taken.
The authors also discussed potential designs for a complementary website for deeper value reflection, which included features like offering multiple ways of reflection, breaking the reflection process down into small steps, including \textit{why} questions etc.

\subsection{Value-Sensitive Design}
Value-sensitive design (VSD) aims to guide ``insightful investigations into technological innovation in ways that foreground the well-being of human beings and the natural world'' \cite{friedman2019value}. Its technology-agnostic nature allowed it to be used in various fields, including artificial intelligence, human-robot interaction, and wastewater treatment \cite{friedman2019value}. Of particular relevance is the toolkit of VSD methods (Table 3.1 in \cite{friedman2019value}). While these methods have been primarily used to design technologies (defined as ``the application of scientific knowledge to solve practical problems'') that reflects the stakeholders' values \cite{friedman2019value}, they could be adapted for reflection on one's own values as well. For example, value tensions could exist not only between different stakeholders, but also within a single individual. We discuss an example research direction inspired by VSD in more detail later (Section \ref{sec:contextVR}).

\subsection{Reflection Frameworks}

Within HCI reflection research, a commonly used framework is Fleck and Fitzpatrick's classification of five types of reflection \cite{fleck2010reflecting}, which includes R0 Description, R1 Reflective Description, R2 Dialogic Reflection, R3 Transformative Reflection and R4 Critical Reflection.
They also discussed various ways that reflection could be supported technologically, e.g., using technology-supported recording for R0 and question-asking for R1, and exploration of alternative perspectives for R2. For R3-4, they suggested that technologies supporting R0-2 could be engaged at a deeper level.
Using this framework, a meta-analysis of personal informatics systems by Cho et al. \cite{cho2022reflection} found that to encourage transformative reflection, future systems could use reflection prompts that are more focused on the \textit{why}'s of recorded user behaviors. In value reflection, the \textit{why} questions can be about values, i.e., how values could have contributed to the recorded user behaviors. As we will discuss later (Section \ref{sec:designaspects}), other technological aspects can also be considered to increase the opportunities for R3-4 to occur during value reflection.

Outside the HCI literature, many works have highlighted the importance of values during reflection. For example, in teaching contexts, being aware of one's own values and their influence on one's teaching approach is crucial during reflection in the teacher preparation process \cite{nolan2008encouraging}. Within adult education research, Mezirow's transformative learning theory is of particular relevance.
The theory describes three types of reflection: content reflection (``reflecting on what we perceive, think, feel, and act'' \cite{taylor2017critical}), process reflection (``reflecting on how we perform the functions of perceiving'' \cite{taylor2017critical}) and premise reflection, which involves ``an examination of the premise or basis of the problem'' and ``engages learners in seeing themselves and the world in a different way'' \cite{cranton2006understanding}. In Mezirow's later research, premise reflection eventually evolved into critical reflection, which is ``about questioning deeply held assumptions about how an individual makes meaning'' of their world \cite{taylor2017critical}.
Another concept from transformative learning theory is the six habits of mind, which are different aspects of how we see ``the world based on our background, experience, culture, and personality'' \cite{cranton2006understanding}.
Based on Mezirow's findings, Cranton \cite{cranton2006understanding} proposed a template to generate reflection questions that can better support critical reflection. As part of this, questions related to the moral-ethical habit of mind involve reflecting on values, and could be used as a reference when designing value reflection processes:
\begin{enumerate}
    \item Content reflection: What are my values?
    \item Process reflection: How have my values formed?
    \item Premise reflection: Why are my values important?
\end{enumerate}

\section{Elements of Value Reflection and Technology's Role}\label{sec:designaspects}

\begin{figure*}[!ht]
    \includegraphics[width=.7\textwidth]{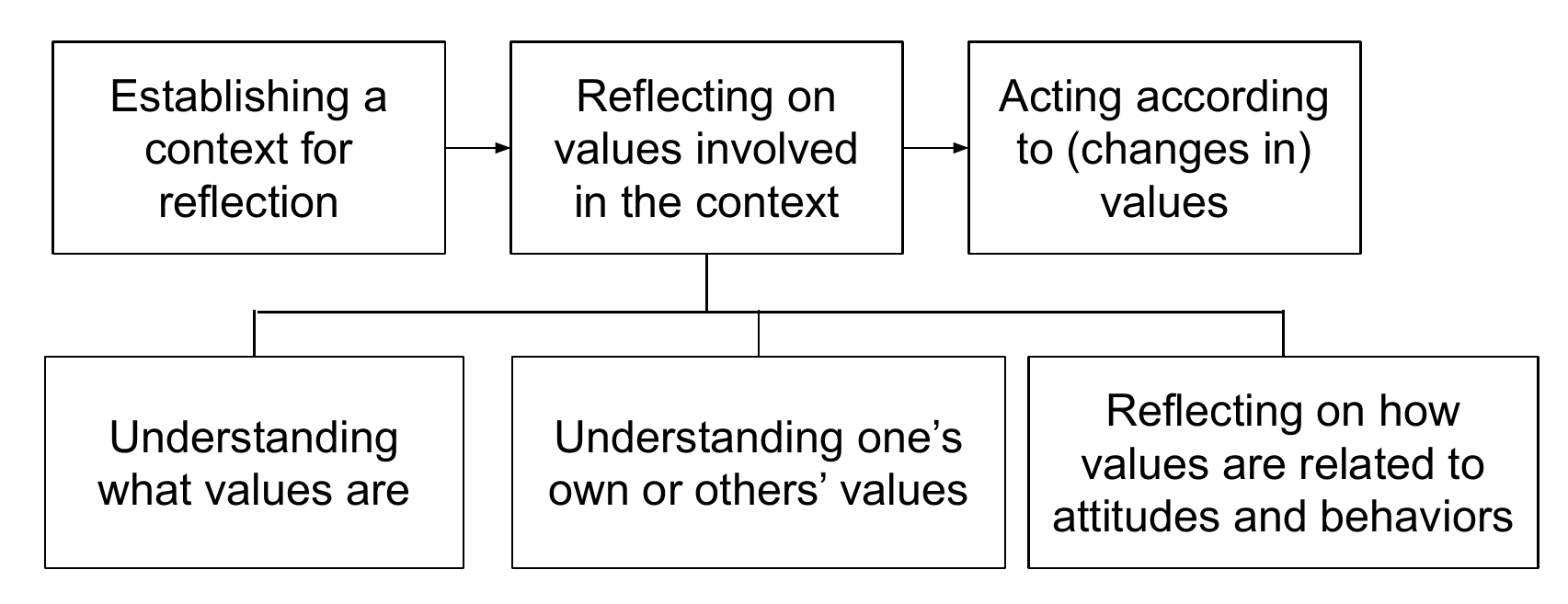}
    \caption{Possible elements that could constitute a value reflection process.}
    \label{fig:possible_elements}
\end{figure*}

\change{In this section, we discuss possible elements that a value reflection process might include.
Since relatively little work has been done on value reflection, these elements are far from determined or fully understood. Instead, we hope that they can serve as a stepping stone for further discussions on how to possibly design for value reflection in various contexts, and the role that technology could play.}

\subsection{Establishing a Context for Reflection}
\change{While having a context to reflect on is nothing unique to value reflection, having a clear context established might be especially important for value reflection.}
As highlighted by Pommeranz et al., values could ``differ in roles and importance to people depending on the situational context in which they guide human behavior or decisions'' \cite{pommeranz2012elicitation}. As such, a well-defined context should be established during value reflection. This can be done either through clear descriptions of the context, or through in-situ reflection processes. As presented in Section \ref{sec:contextVR}, the context of a task-specific value reflection process performed during the task itself is likely clearer. 

\subsection{Context-Specific Value Reflection}
\change{Once the context has been established, the next step is to reflect on the values involved. The target of reflection could vary widely, be it an artifact, past events or personal experiences \cite{leecompassion,pommeranz2012elicitation}. Moreover, the actual method through which reflection is done could also take various forms, e.g., verbal prompts, daily diary entries, methods inspired by existing frameworks like VSD. 
Importantly, value reflection prompts should be designed to be grounded within a specific context, since the meaningfulness of over-generalized prompts (e.g., asking ``what values are important to you?'' without providing a context prior) might be limited at best. 
Often, the desired outcome of this step might be a change in personal values, or a person's understanding of values.}

As such, during this step, it might be important for technologies to facilitate people in understanding what values are, their own values, and the values of others. As Roccas et al. highlighted, values ``must be represented cognitively in ways that enable people to communicate about them'' (Pg. 62 of \cite{maio2016psychology}). This statement mirrors work by Huldtgren et al. \cite{huldtgren2014designing}, who found that participants had a hard time understanding the values they were given when asking them to reflect on those values. Technology could help with this by helping people scaffold their process of learning what values mean. Findings from existing research on scaffolding in education (e.g., \cite{gibbons2002scaffolding}), navigating the ladder of abstraction (e.g., \cite{seabury1991critical}), and particularly the role of technology in this process \cite{sharma2007scaffolding} might be especially relevant.

Moreover, as researchers have noted, it is not unlikely for strong emotional reactions to be elicited when an individual's values are challenged \cite{maio2016psychology}. As such, when bringing in values during reflection, it is important for researchers to make sure that the process handles participants' emotions with care. Several types of technologies could aid in this process. For example, emotion detection and sentiment analysis technology could help to recognize emotions, and employ emotion regulation technology whenever appropriate, all scaffolded into the value reflection process.
Importantly, value reflection processes should not try to completely avoid situations that might result in responses or reactions that are more emotional, since it could be an essential part of undergoing a long-term change in values, e.g., a disorienting dilemma is the very first step of transformative learning \cite{Kitchenham2008}. However, the opposite extreme of designing for extremely emotional responses is irresponsible and could result in harm. As such, personalization of reflection processes that allow people to open up and perform deeper reflections on their values, while not becoming too emotional, is crucial. 
How can technology help with this balancing act while respecting the users' privacy, and without taking advantage of their trust? 
How can technology be designed to provide users with digital or virtual spaces in which they are comfortable opening up and exploring their inner vulnerabilities (e.g., through role-playing \cite{harrell2018chimeria})?


Understanding the effects of values on other constructs, like attitudes and behaviors, might also be crucial.
When administering a value reflection process, Huldtgren et al. found that participants had difficulties drawing clear connections between their values and experiences or thoughts \cite{huldtgren2014designing}. Technology could aid in this through novel visualizations and interactions as part of the reflection process. For example, by having a graph-like visualization where users can explore how their actions, thoughts or experiences could be connected to their values. By encouraging explicit considerations of how these constructs are linked to how they prioritize values, users could learn more clearly the effects of values in the specific context that is reflected upon. Moreover, technology could also help with identifying actions and thoughts that might be in conflict with values they prioritize. Exposing such conflicts could help people better align their actions, thoughts and values, either through changing their values, or changing their behaviors and thoughts. 

\subsection{Acting According to (Changes in) Values}
The final component of a value reflection process, after understanding values (and possibly value changes) and their effects, could be to engender an actual long-term change in their way of thinking and/or behavior. Technology could help by encouraging people to imagine how might they change their behaviors or attitudes in the future, for instance, by simulating possible future scenarios while keeping their values in mind. 
Existing research on habit-forming technologies (e.g., \cite{10.1145/3196830}) could also be of particular relevance here.

\section{Example Contexts for value reflection}\label{sec:examples}
To provide more concrete examples of potential avenues for value reflection \change{and how previously discussed topics could take shape}, here is a quick discussion of our past (Section \ref{sec:climatechange}) and ongoing (Section \ref{sec:contextVR}) research efforts that investigate value reflection in two different contexts.

\subsection{Climate Change Digital Media}\label{sec:climatechange}
Digital media has become an important part of the daily lives of many. YouTube, for example, has 1.7 billion unique monthly users who are on YouTube for a daily average of 19 minutes, streaming 694,000 hours of videos every minute \cite{youtubestats}. While a large amount of digital content on various global and societal issues exists, the expected translation from media consumption to compassionate and prosocial behaviors has largely not been realized (e.g., media on human suffering \cite{Hoskins_2020}). In this study \cite{leecompassion}, we wanted to investigate if asking people reflection questions before and after watching a climate change documentary from YouTube can better motivate people to be more compassionate, and adopt environmentally friendly (i.e., prosocial) behaviors. Specifically, before watching the video, participants in the reflection conditions were asked to describe their views and values on climate change, and what factors helped form their views and values. Interestingly, a few participants talked about how their digital media use played a big role, e.g., following Instagram creators who primarily create content to promote environmental awareness. This observation aligns with prior research showing that social media interactions are an important social context that can influence people's values, and consequently, their attitudes and actions \cite{maio2016psychology}.
After watching the video, they were asked to reflect on specific moments from the video (that they chose), and how their values and perspectives might have changed after watching those moments. Finally, they were asked to think about how these changes might affect their actions in the future. After two weeks, all participants were asked if their behaviors have changed in the past two weeks. We found that participants in the reflection conditions who engaged with the reflection process in the first session (i.e., they performed deeper levels of reflection during the first session) reported significantly more prosocial behavioral changes. Examples include using more environmentally friendly modes of transportation and consumer behaviors (e.g., buying less meat, stopped using plastic bags). 
Even though reflection questions \change{(which were created based on Cranton's template \cite{cranton2006understanding})} were asked verbally by a researcher in this study, findings provide support for potential benefits of a more technologically-supported value reflection process in digital media contexts \cite{leecompassion}.


\subsection{Collaborative VR Tasks}\label{sec:contextVR}
We will now present briefly our current research on designing value reflection in the specific context of collaboratively designing virtual spaces
in virtual reality (VR).
In general, VR collaboration has seen increasing use cases (e.g., construction \cite{Podkosova_2022,thewild}, health \cite{realmedical}, education \cite{Wang_2021}) and is expected to become more commonplace with growing VR adoption (e.g., collaboratively designing virtual environments in social VR applications). 
Collaborations often fail when stakeholders' values are not sufficiently taken into account.
However, while stakeholder value elicitation and negotiation methods exist to mitigate this \cite{grunbacher2006stakeholder}, stakeholders might not have a good understanding of their own values.

Inspired by VSD methods like value sketching and the value sensitive action-reflection model \cite{friedman2019value}, we are interested in designing incremental prompts used in contexts where the designers themselves are also stakeholders.
We plan for these prompts to be used in the design process to incrementally encourage deeper reflection on designers' own values in-situ (i.e., while in VR) and a better alignment between their values and design decisions.
Taking a task of designing virtual towns as an example, prompts could encourage reflection by introducing more diverse perspectives through automatic identification of alignments and conflicts in values represented by different buildings and their spatial placements.
These prompts could also be easily extended to highlight value similarities and differences between a designer and other stakeholders, and encourage reflection on compromises and the balance between self and others.

\section{Conclusion}
While factors other than values could also predict an individual's behaviors (e.g., external constraints, other individuals, social norms \cite{maio2016psychology}), values play an important role in shaping our actions and attitudes nonetheless. In this position paper, we briefly discussed why HCI researchers should consider the inclusion of context-specific values as reflection targets when designing reflection processes. We also provided a high-level overview of some relevant work and design aspects that could inspire and aid researchers in the process of designing value reflection processes.

\bibliographystyle{ACM-Reference-Format}
\bibliography{main}


\end{document}